\documentclass[preprintnumbers,amsmath,amssymbm,prd]{revtex4}
\usepackage{epsfig}
\usepackage{graphicx}
\usepackage{amssymb}

\begin{document}
\title{Spinning Kerr black holes with stationary massive scalar clouds: The large-coupling regime}
\author{Shahar Hod}
\affiliation{The Ruppin Academic Center, Emeq Hefer 40250, Israel}
\affiliation{ } \affiliation{The Hadassah Institute, Jerusalem
91010, Israel}
\date{\today}

\begin{abstract}
We study analytically the Klein-Gordon wave equation for stationary massive scalar fields linearly coupled to spinning 
Kerr black holes. In particular, using the WKB approximation, we derive a compact formula for the discrete spectrum of scalar field masses which characterize the stationary composed Kerr-black-hole-massive-scalar-field configurations in the large-coupling regime $M\mu\gg1$ (here $M$ and $\mu$ are respectively the mass of the central black hole and the proper mass of the scalar field). We confirm our analytically derived formula for the Kerr-scalar-field mass spectrum with 
numerical data that recently appeared in the literature. 
\end{abstract}
\bigskip
\maketitle

\section {Introduction}

The complex interactions between fundamental matter fields and general relativistic black holes have been the focus of intense research efforts during the last five decades (see \cite{Chan,Zel,PressTeu2,Bekc,Hodkr,Herkr,Notemas,CarDias} and references therein). Especially interesting are the physical properties of astrophysically realistic spinning black holes coupled to bosonic matter fields. As explicitly demonstrated recently \cite{Hodkr,Herkr}, these integer-spin fields can form non-trivial spatially regular {\it stationary} matter configurations in rotating curved black-hole spacetimes. 

In particular, recent analytical \cite{Hodkr} and numerical \cite{Herkr} studies of the stationary sector of the coupled Einstein-Klein-Gordon field equations have revealed the intriguing fact that, for a given spinning black hole of mass $M$ and angular momentum per unit mass $a$, there exists a {\it discrete} spectrum of resonant field masses, $\{M\mu(a/M,m;n)\}^{n=\infty}_{n=0}$ \cite{Noteunit,Notedim,Notemn,Notesmp}, which characterize a family of stationary spatially regular scalar configurations that can be supported in the external region of the corresponding rotating black-hole spacetime. 

Interestingly, these stationary bound-state bosonic matter configurations are
characterized by proper azimuthal frequencies which are in resonance with (that is, are multiple integers of) the angular frequency of the supporting central black hole \cite{Hodkr,Herkr,Notecrit}:
\begin{equation}\label{Eq1}
\omega_{\text{field}}=\omega_{\text{c}}\equiv
m\Omega_{\text{H}}\  ,
\end{equation}
where the integer $m$ is the azimuthal harmonic index of the scalar mode and \cite{Chan,Kerr,Noterp}
\begin{equation}\label{Eq2}
\Omega_{\text{H}}={{a}\over{r^2_++a^2}}\
\end{equation}
is the angular velocity which characterizes the outer horizon of the spinning Kerr black hole. 

For a scalar field of proper mass $\mu$, the resonant field frequency (\ref{Eq1}) which, for a given value of the 
azimuthal harmonic index $m$, characterizes a stationary bound-state scalar configuration in the Kerr black-hole spacetime, is also bounded from above by the inequality
\cite{Hodkr,Herkr}
\begin{equation}\label{Eq3}
\omega^2_{\text{field}}<\mu^2\  .
\end{equation}
The characteristic relation (\ref{Eq3}) guarantees that the massive scalar configuration is spatially bounded to the black hole. In particular, low frequency stationary scalar modes in the regime (\ref{Eq3}) are captured inside an effective potential well which characterizes the composed black-hole-massive-scalar-field system [see Eq. (\ref{Eq18}) below]. Thus, matter configurations made of massive scalar fields in the regime (\ref{Eq3}) cannot radiate their energies to asymptotic infinity \cite{Hodkr,Herkr,Notediv}.

The discrete spectrum of resonant field masses $\{M\mu(a/M,m)\}$, which characterizes the stationary bound-state scalar configurations in the spinning Kerr black-hole spacetime, has been computed {\it numerically} in \cite{Herkr}. For rapidly-rotating (near-extremal) black holes in the regime $a/M\to 1$, the characteristic mass spectrum of the composed black-hole-linearized-massive-scalar-field configurations has been studied {\it analytically} in \cite{Hodkr}. To the best of our knowledge, the characteristic resonant spectrum $\{M\mu(a/M,m)\}$ of the stationary bound-state massive scalar configurations has not been determined analytically thus far for {\it generic} \cite{Notegen} values $a/M\in(0,1)$ of the dimensionless rotation parameter which characterizes the central supporting black holes. 

The main goal of the present paper is to derive a compact {\it analytical} formula for the discrete spectrum $\{M\mu(a/M,m)\}$ of field masses which characterize the stationary bound-state scalar configurations in {\it generic} \cite{Notegen} Kerr black-hole spacetimes. To this end, we shall study analytically the Klein-Gordon wave equation for stationary massive scalar fields linearly coupled to rotating Kerr black-hole spacetimes. In particular, using the WKB approximation, we shall derive below a compact analytical formula for the discrete spectrum $\{M\mu(a/M,m)\}$ of scalar field masses which characterize the stationary bound-state Kerr-black-hole-massive-scalar-field configurations in the eikonal large-mass $M\mu\gg1$ regime. Most importantly, this analytically derived formula [see Eq. (\ref{Eq27}) below] for the scalar resonance spectrum would be valid in the {\it entire} physical range $a/M\in(0,1)$ of the black-hole rotation parameter.

\section{Description of the system}

We shall analyze the properties of a physical system which is composed of a stationary scalar field $\Psi$ 
of mass $\mu$ and proper frequency $\omega_{\text{field}}=m\Omega_{\text{H}}$ [see Eqs. (\ref{Eq1}) and (\ref{Eq2})] which
is linearly coupled to a central rotating Kerr black hole of mass $M$ and angular momentum per unit mass $a$. 
Using the Boyer-Lindquist coordinates $(t,r,\theta,\phi)$, 
one can describe the spinning black-hole
spacetime by the line element \cite{Chan,Kerr}
\begin{eqnarray}\label{Eq4}
ds^2=-{{\Delta}\over{\rho^2}}(dt-a\sin^2\theta
d\phi)^2+{{\rho^2}\over{\Delta}}dr^2+\rho^2
d\theta^2+{{\sin^2\theta}\over{\rho^2}}\big[a
dt-(r^2+a^2)d\phi\big]^2\  ,
\end{eqnarray}
where 
\begin{equation}\label{Eq5}
\Delta\equiv r^2-2Mr+a^2\ \ \ ; \ \ \ \rho^2\equiv
r^2+a^2\cos^2\theta\  .
\end{equation}
The zeros of $\Delta$,
\begin{equation}\label{Eq6}
r_{\pm}=M\pm\sqrt{M^2-a^2}\  ,
\end{equation}
determine the horizon radii of the Kerr black-hole spacetime.

The spatial and temporal properties of a linearized massive scalar field in the curved black-hole spacetime are determined by the familiar 
Klein-Gordon wave equation
\begin{equation}\label{Eq7}
(\nabla^{\nu}\nabla_{\nu}-\mu^2)\Psi=0\  .
\end{equation}
Substituting the metric components of the Kerr black-hole line element (\ref{Eq4}) into the Klein-Gordon wave equation
(\ref{Eq7}) and using the scalar field decomposition \cite{Noteanz}
\begin{equation}\label{Eq8}
\Psi(t,r;\omega,\theta,\phi)=\sum_{l,m}e^{im\phi}{S_{lm}}(\theta;m,a\sqrt{\mu^2-\omega^2})
{R_{lm}}(r;M,a,\mu,\omega)e^{-i\omega t}\  ,
\end{equation}
one obtains the characteristic ordinary differential equation (known as the spheroidal angular equation) \cite{Stro,Heun,Fiz1,Teuk,Abram,Hodasy,Barma,Hodpp}
\begin{eqnarray}\label{Eq9}
{1\over {\sin\theta}}{{d}\over{d\theta}}\Big(\sin\theta {{d
S_{lm}}\over{d\theta}}\Big)
+\Big[K_{lm}+a^2(\mu^2-\omega^2)\sin^2\theta-{{m^2}\over{\sin^2\theta}}\Big]S_{lm}=0\
\end{eqnarray}
for angular part $S_{lm}$ of the scalar eigenfunction $\Psi$ \cite{Notespd}. 
Spheroidal harmonic eigenfunctions $\{S_{lm}\}$ which are regular at the angular poles
$\theta=0$ and $\theta=\pi$ are known to be characterized by a discrete set
$\{K_{lm}\}$ of angular eigenvalues \cite{Stro,Heun,Fiz1,Teuk,Abram,Hodasy,Barma,Hodpp}. In particular, it has recently been proved 
that, in the eikonal ($|m|\gg1$) regime, the angular eigenvalues are given by the asymptotic large-$m$ analytical 
formula \cite{Hodpp,Notebpo,Noteklm}
\begin{equation}\label{Eq10}
K_{lm}=m^2-a^2(\mu^2-\omega^2)+[2(l-|m|)+1]\sqrt{m^2+a^2(\mu^2-\omega^2)}+O(1)\  .
\end{equation}

Substituting the scalar field decomposition (\ref{Eq8}) into (\ref{Eq7}), and using the line element (\ref{Eq4}) which characterizes the spinning Kerr black-hole spacetime, one finds the characteristic ordinary differential equation \cite{Teuk,Stro}
\begin{equation}\label{Eq11}
\Delta{{d} \over{dr}}\Big(\Delta{{d R_{lm}
}\over{dr}}\Big)+\Big\{[\omega(r^2+a^2)-ma]^2
+\Delta[2ma\omega-\mu^2(r^2+a^2)-K_{lm}]\Big\}R_{lm}=0\ 
\end{equation}
for the radial part $R_{lm}$ of the scalar eigenfunction $\Psi$, where the angular eigenvalues $\{K_{lm}\}$ that explicitly appear in the radial equation (\ref{Eq11}) should be determined from the spheroidal differential equation (\ref{Eq9}) [see, in particular, the asymptotic angular spectrum (\ref{Eq10}) \cite{Hodpp}]. Physically acceptable solutions of the scalar radial equation (\ref{Eq11}) are characterized by purely ingoing waves (as measured by a comoving observer) at the black-hole outer horizon \cite{Notemas,Hodkr,Herkr}:
\begin{equation}\label{Eq12}
R_{lm} \sim e^{-i(\omega-m\Omega_{\text{H}})y}\ \ \ \text{ for }\ \
\ r\rightarrow r_+\ \ (y\rightarrow -\infty)\  ,
\end{equation}
where the radial coordinate $y$, to be used below, is defined by the differential relation
$dy=(r^2/\Delta)dr$ [see Eq. (\ref{Eq15}) below]. In addition, the spatially regular bound-state (finite-mass) configurations of the massive scalar fields in the black-hole spacetime are characterized by asymptotically decaying radial eigenfunctions at spatial infinity \cite{Notemas,Hodkr,Herkr}:
\begin{equation}\label{Eq13}
R_{lm} \sim {1\over r}e^{-\sqrt{\mu^2-\omega^2}y}\ \ \ \text{ for }\
\ \ r\rightarrow\infty\ \ (y\rightarrow \infty)\  .
\end{equation}

The physically motivated boundary conditions (\ref{Eq12}) and (\ref{Eq13}) of the radial scalar equation (\ref{Eq11}), together with the resonance condition (\ref{Eq1}), determine the discrete spectrum $\{M\mu(m,a/M;n)\}$ of scalar field masses which characterize the stationary composed Kerr-black-hole-linearized-massive-scalar-field configurations. In the next section we shall use {\it analytical} techniques in order to determine the characteristic Kerr-scalar-field resonance spectrum in the large mass ($M\mu\gg1$) regime. 

\section{The resonance mass spectrum of the composed Kerr-black-hole-massive-scalar-field configurations}

In the present section we shall derive a compact analytical formula for the discrete spectrum $\{M\mu(m,a/M;n)\}$ 
of scalar field masses which characterize the composed stationary Kerr-black-hole-massive-scalar-field configurations in the eikonal large-mass regime. The trick is to transform the radial differential equation (\ref{Eq11}) of the scalar field into the familiar form of a Schr\"odinger-like radial differential equation and then to perform a standard WKB analysis.

Defining the new radial eigenfunction
\begin{equation}\label{Eq14}
\psi=rR\
\end{equation}
and using the coordinate transformation \cite{Notemap}
\begin{equation}\label{Eq15}
dy={{r^2}\over{\Delta}}dr\  ,
\end{equation}
one obtains from (\ref{Eq11}) the
Schr\"odinger-like ordinary differential equation
\begin{equation}\label{Eq16}
{{d^2\psi}\over{dy^2}}-V(y)\psi=0\  ,
\end{equation}
with the effective radial potential 
\begin{equation}\label{Eq17}
V=V(r;\omega,M,a,\mu,l,m)={{2\Delta}\over{r^6}}(Mr-a^2)+{{\Delta}\over{r^4}}
[K_{lm}-2ma\omega+\mu^2(r^2+a^2)]-{{1}\over{r^4}}[\omega(r^2+a^2)-ma]^2\
.
\end{equation}

Substituting into (\ref{Eq17}) the resonant oscillation frequency (\ref{Eq1}), which characterizes the stationary massive 
scalar field configurations in the spinning Kerr black-hole spacetime, and using the asymptotic large-$m$ analytical 
formula (\ref{Eq10}) for the characteristic eigenvalues of the angular differential equation (\ref{Eq9}), one finds the expression 
\begin{equation}\label{Eq18}
V(r)=m^2\cdot \Big\{{{r-r_+}\over{r^3(r^2_++a^2)^2}}\big[\beta
a^2r^2-a^2(2M+\beta
r_-)r+2Mr^3_+\big]+{{1}\over{m}}\cdot{{\Delta}\over{r^4}}(2k+1)\sqrt{1+\beta{{a^4}\over{(r^2_++a^2)^2}}}+O(m^{-2})\Big\}\
\end{equation}
for the effective radial potential which characterizes the composed 
stationary Kerr-black-hole-massive-scalar-field configurations in the eikonal $m\gg1$ regime \cite{Notelarmu}, where $k\equiv l-m\ll m$ \cite{Noteklm}. Here we have used the dimensionless physical parameter $\beta$ which relates the proper mass of the stationary scalar field to its resonant oscillation frequency \cite{Notebet}:
\begin{equation}\label{Eq19}
\mu^2=(1+\beta)\cdot\omega^2_{\text{c}}\  .
\end{equation}

As we shall now show, the characteristic Schr\"odinger-like differential equation (\ref{Eq16}) 
is amenable to a standard WKB analysis \cite{WKB1,WKB2,WKB3,Will,Hodalm}. 
In particular, one finds that the effective radial potential (\ref{Eq18}) of the stationary composed Kerr-massive-scalar-field configurations 
is characterized by the presence of a {\it binding} potential well outside the black-hole horizon \cite{Hodbmu}. As shown in \cite{WKB1,WKB2,Hodalm}, the WKB resonance condition which
characterizes the bound-state resonances of the standard Schr\"odinger-like ordinary differential equation (\ref{Eq16})
in the eikonal large-$m$ regime is given by
\begin{equation}\label{Eq20}
{V_{\text{min}}\over{\sqrt{2V^{''}_{\text{min}}}}}=-(n+{1\over 2})\ \ \ ; \ \ \ n=0,1,2,...\   ,
\end{equation}
where a prime denotes a spatial derivative of the effective radial potential (\ref{Eq18}) with respect to the radial coordinate $y$ [see Eq. (\ref{Eq15})] of the Schr\"odinger-like differential equation. The subscript ``min" in (\ref{Eq20}) means that the effective radial potential and its spatial derivatives are evaluated at the radial coordinate $r^{\text{min}}$ which characterizes the minimum point of the effective binding potential (\ref{Eq18}) \cite{Notemin}.

The WKB resonance equation (\ref{Eq20}) with the (rather cumbersome) radial potential (\ref{Eq18}) can be solved analytically using an iteration scheme (that is, using an expansion in inverse powers of $m$). Keeping only the leading order $O(m^2)$ term of the effective radial potential (\ref{Eq18}), one finds from (\ref{Eq20}) the leading-order solution
\begin{equation}\label{Eq21}
r^{\text{min}}_0=r_+\cdot {{\gamma[1-\sqrt{1-\gamma^2}]}\over{2-\gamma^2-2\sqrt{1-\gamma^2}}}\
\end{equation}
with
\begin{equation}\label{Eq22}
\beta_0={{(1+\gamma)(2-\gamma^2-2\sqrt{1-\gamma^2})}\over{\gamma^3}}\  ,
\end{equation}
where
\begin{equation}\label{Eq23}
\gamma\equiv {{r_-}\over{r_+}}\
\end{equation}
is the dimensionless ratio between the horizon radii [see Eq. (\ref{Eq6})] of the rotating Kerr black-hole spacetime. 

Next, substituting
\begin{equation}\label{Eq24}
r^{\text{min}}=r^{\text{min}}_0\cdot\big[1+{{r_1}\over{m}}+O(m^{-2})\big]\
\end{equation}
and
\begin{equation}\label{Eq25}
\beta=\beta_0\cdot\big[1+{{\beta_1}\over{m}}+O(m^{-2})\big]\  ,
\end{equation}
and keeping terms of orders $O(m^2)$ and $O(m)$ in the effective radial potential (\ref{Eq18}), one finds from (\ref{Eq20}) after 
some lengthy algebra \cite{Noter1}
\begin{equation}\label{Eq26}
\beta_1=-{{(2-\gamma^2-2\sqrt{1-\gamma^2})\sqrt{1+{{1}\over{\gamma}}}}\over{(1-\sqrt{1-\gamma^2})^2}}\cdot
\Big[\sqrt{-1+\gamma+\sqrt{1-\gamma^2}}\cdot(1+2n)+\sqrt{2+\gamma-2\sqrt{1-\gamma^2}}\cdot(1+2k)\Big]\  .
\end{equation}

Finally, taking cognizance of Eqs. (\ref{Eq1}), (\ref{Eq2}), (\ref{Eq19}), (\ref{Eq22}), (\ref{Eq23}), (\ref{Eq25}), and (\ref{Eq26}), one finds the analytical formula 
\begin{equation}\label{Eq27}
M\mu=m\cdot [\Gamma_0+{{\Gamma_1}\over{m}}+O(m^{-2})]\
\end{equation}
for the discrete spectrum of scalar field masses which characterize the stationary composed Kerr-black-hole-massive-scalar-field configurations in the eikonal large-mass ($M\mu\gg1$) regime, where
\begin{equation}\label{Eq28}
\Gamma_0={{\sqrt{2(1+\gamma)(1-\sqrt{1-\gamma^2})-\gamma^2}}\over{2\gamma}}\  
\end{equation}
and \cite{Noteg1}
\begin{equation}\label{Eq29}
\Gamma_1=-{{(1+{{1}\over{\gamma}})^{3/2}(2-\gamma^2-2\sqrt{1-\gamma^2})^2\Big[\sqrt{-1+\gamma+\sqrt{1-\gamma^2}}\cdot(1+2n)+\sqrt{2+\gamma-2\sqrt{1-\gamma^2}}\cdot(1+2k)\Big]}\over{4(1-\sqrt{1-\gamma^2})^2\sqrt{2(1+\gamma)(1-\sqrt{1-\gamma^2})-\gamma^2}}}\  .
\end{equation}

\section{Numerical confirmation}

It is of physical interest to test the accuracy of the analytically
derived formula (\ref{Eq27}) for the discrete mass spectrum of the composed stationary 
Kerr-black-hole-massive-scalar-field configurations. The corresponding scalar field masses of the stationary configurations 
were computed numerically in the very interesting work of Herdeiro and Radu \cite{Herkr}.  
In Table \ref{Table1} we display the dimensionless ratio
${{\mu_{\text{numerical}}}/{\mu_{\text{wkb}}}}$ for the fundamental ($n=0$) resonant mode with $l=m=10$, where
$\{\mu_{\text{numerical}}(a/M)\}$ are the {\it numerically} computed \cite{Herkr}
field masses of the stationary scalar configurations and
$\{\mu_{\text{wkb}}(a/M)\}$ are the {\it analytically} derived masses of the stationary bound-state scalar fields as given by the WKB expression (\ref{Eq27}). From the data presented in Table \ref{Table1} one finds that the agreement between the numerical data \cite{Herkr} and the
analytically derived formula (\ref{Eq27}) for the resonant masses of the stationary black-hole-scalar-field configurations is generally better than $1\%$ \cite{Notenex}.


\begin{table}[htbp]
\centering
\begin{tabular}{|c|c|}
\hline \ \ $s\equiv {{a}/{M}}$\ \ & \ \
${{\mu(l=m=10)}\over{\mu_{\text{wkb}}}}$\ \ \\
\hline
\ \ 0.1\ \ & \ \ 0.99946\ \ \ \\
\ \ 0.2\ \ & \ \ 0.99993\ \ \ \\
\ \ 0.3\ \ & \ \ 1.00009\ \ \ \\
\ \ 0.4\ \ & \ \ 1.00017\ \ \ \\
\ \ 0.5\ \ & \ \ 0.99974\ \ \ \\
\ \ 0.6\ \ & \ \ 1.00037\ \ \ \\
\ \ 0.7\ \ & \ \ 1.00076\ \ \ \\
\ \ 0.8\ \ & \ \ 1.00134\ \ \ \\
\ \ 0.9\ \ & \ \ 1.00266\ \ \ \\
\ \ 0.95\ \ & \ \ 1.00399\ \ \ \\
\ \ 0.99\ \ & \ \ 1.00750\ \ \ \\
\ \ 0.999\ \ & \ \ 1.01219\ \ \ \\
\hline
\end{tabular}
\caption{Stationary massive scalar clouds linearly coupled to spinning Kerr black holes. We display the dimensionless ratio
${{\mu_{\text{numerical}}}/{\mu_{\text{wkb}}}}$, where
$\{\mu_{\text{numerical}}(a/M)\}$ are the {\it numerically} computed \cite{Herkr}
field masses of the stationary scalar fields and
$\{\mu_{\text{wkb}}(a/M)\}$ are the {\it analytically} derived scalar field masses as given by the WKB expression (\ref{Eq27}). 
The data presented are for the fundamental ($n=0$) resonant mode with 
$l=m=10$. One finds that the agreement between the numerical data and the
analytical formula (\ref{Eq27}) is generally better than $1\%$.}
\label{Table1}
\end{table}

\section{Summary}

The intriguing physical mechanism of rotational superradiance
\cite{Zel,PressTeu2} allows spinning Kerr black holes to support stationary bound-state matter configurations made of 
massive scalar fields in their exterior regions. The physical properties of these stationary composed black-hole-field configurations 
have been studied extensively in recent years \cite{Hodkr,Herkr}. Interestingly, it has been shown \cite{Hodkr,Herkr} that, for 
a given set $\{M,a,m\}$ of the black-hole-field physical parameters, there exists a {\it discrete} spectrum $\{M\mu(a/M,m;n)\}$ of field masses which characterize the composed stationary Kerr-black-hole-massive-scalar-field configurations. In particular, for near-extremal (rapidly-rotating) Kerr black holes, the characteristic discrete mass spectrum of the composed black-hole-linearized-massive-scalar-field configurations has been studied {\it analytically} in \cite{Hodkr}.

In the present paper we have studied analytically the Klein-Gordon wave equation for stationary massive scalar fields linearly coupled to generic \cite{Notegen} spinning Kerr black holes. In particular, using the WKB approximation, we have derived an explicit analytical formula [see Eqs. (\ref{Eq27})-(\ref{Eq29})] for the discrete spectrum of scalar field masses which characterize the stationary composed Kerr-black-hole-massive-scalar-field configurations in the eikonal large-mass $M\mu\gg1$ regime. Interestingly, the {\it analytically} derived formula (\ref{Eq27}) for the discrete mass spectrum of the composed Kerr-scalar-field configurations was shown to agree remarkably well with direct {\it numerical} computations \cite{Herkr} of the corresponding black-hole-massive-scalar-field resonances. 

Finally, it is worth noting that, to the best of our knowledge, the analytically derived expression (\ref{Eq27}), which is valid in the entire physical range $a/M\in(0,1)$ of the black-hole rotation parameter \cite{Noteacc}, provides the first analytical formula for the discrete mass spectrum of the stationary bound-state linearized massive scalar field configurations in the spacetimes of {\it generic} \cite{Notegen} spinning Kerr black holes.

\bigskip
\noindent
{\bf ACKNOWLEDGMENTS}
\bigskip

This research is supported by the Carmel Science Foundation. I would
like to thank C. A. R. Herdeiro and E. Radu for sharing with me
their numerical data. I thank Yael Oren, Arbel M. Ongo, Ayelet B.
Lata, and Alona B. Tea for stimulating discussions.



\begin{thebibliography}{99}
\bibitem{Noteunit} We shall use natural units in which $G=c=\hbar=1$.
\bibitem{Notedim} Note that the proper mass $\mu$, which characterizes the scalar field mode, stands for $\mu/\hbar$. Hence,
this physical parameter of the matter field has the dimensions of $($length$)^{-1}$.
\bibitem{Notemn} Here the dimensionless integer parameters $m$ and $n$ are respectively the azimuthal harmonic index and the resonance parameter of the stationary bound-state massive scalar field mode [see Eqs. (\ref{Eq8}) and (\ref{Eq20}) below].
\bibitem{Notesmp} We shall assume $a>0$ and $m>0$ for the black-hole-field physical parameters without loss of
generality.
\bibitem{Notecrit} It is worth noting that, for a given value of the azimuthal harmonic index $m$ of the scalar mode, the characteristic proper frequency $\omega_{\text{field}}=m\Omega_{\text{H}}$ of the stationary field configuration coincides with the threshold (critical) frequency $\omega_{\text{c}}$ for the superradiant scattering phenomenon of bosonic fields in the spinning Kerr black-hole spacetime \cite{Zel,PressTeu2}.
\bibitem{Noterp} Here $r_+$ is the radius of the outer black-hole horizon [see Eq. (\ref{Eq6}) below].
\bibitem{Noteanz} Here the integer $l\geq|m|$ is the spheroidal 
harmonic index of the bound-state scalar field mode.
\bibitem{Notespd} The angular eigenfunctions $\{S_{lm}\}$ are the familiar spheroidal harmonics.
\bibitem{Notebpo} As explicitly proved in \cite{Hodpp}, the large-$m$ asymptotic formula (\ref{Eq10}) is valid in 
the regime $-a^2(\mu^2-\omega^2)<m^2$. Taking cognizance of the inequality $\omega^2<\mu^2$ which characterizes 
the bound-state massive scalar resonances in the black-hole spacetime [see (\ref{Eq3})], one finds that the requirement $-a^2(\mu^2-\omega^2)<m^2$ is trivially satisfied.
\bibitem{Noteklm} Note that, as explicitly proved in \cite{Hodpp}, the large-$m$ analytical formula (\ref{Eq10}) is valid in the $l-|m|\ll \sqrt{m^2+a^2(\mu^2-\omega^2)}$ regime.
\bibitem{Notemap} It is worth noting that the radial range $r\in [r_+,\infty]$ is mapped into the radial range $y\in [-\infty,+\infty]$ by the radial differential relation (\ref{Eq15}).
\bibitem{Notelarmu} Note that the eikonal large-$m$ regime also corresponds to the large-mass $M\mu\gg1$ regime [see Eq. (\ref{Eq27}) below].
\bibitem{Notebet} Note that the characteristic inequality (\ref{Eq3}) for the bound-state resonances of the massive scalar fields in the black-hole spacetime implies that $\beta>0$ .
\bibitem{Notemin} That is, the radial coordinate $r^{\text{min}}$ is defined by the relation $V^{'}[r=r^{\text{min}}(y)]=0$ with $V^{''}[r=r^{\text{min}}(y)]>0$.
\bibitem{Noter1} The explicit expression of the dimensionless expansion coefficient  $r_1$ is rather cumbersome and is not important 
for our purposes.
\bibitem{Noteg1} Note that $\Gamma_1<0$, which implies $M\mu<m\Gamma_0$ [see Eq. (\ref{Eq27})] in the asymptotic $m\to\infty$ regime \cite{Hodbmu}.
\bibitem{Noteacc} It is worth emphasizing again that, as we explicitly shown in this paper (see the data presented in Table \ref{Table1}), the analytical formula (\ref{Eq27}) is not only valid in the entire physical regime $a/M\in(0,1)$ of the supporting black hole, but it also agrees remarkably well with the exact (numerically computed \cite{Herkr}) proper masses which characterize the stationary bound-state scalar fields in the spinning Kerr black-hole spacetime.
\end{thebibliography}
\end{document}